# High-pressure hybrid materials that can store hydrogen in table salt

Feng Peng,[1,2] Yanming Ma,[3] and Maosheng Miao[2,4]*

[1]*College of Physics and Electronic Information, Luoyang Normal University, Luoyang 471022, China*
[2]*Department of Chemistry and Biochemistry, California State University Northridge, Northridge, CA, 91330-8262, USA*
[3] *State Key Laboratory of Superhard Materials & Innovation Center for Computational Physics Methods and Software, College of Physics, Jilin University, Changchun 130012, China*
[4] *Beijing Computational Science Research Centre, Beijing 100193, China*

**Abstract**

We demonstrate in this paper that high pressure can promote the reactions between the ionic compounds and $H_2$ molecules and form thermodynamically stable hybrid compounds. Using crystal structure search method based on first principles calculations and particle swarm optimization algorithm, we show that many alkali halides XY (X=Na, K, Rb, Cs; Y=Cl, Br, I) can form stable hybrid compounds with various $H_2$ compositions under different pressures that could be as low as 0.8 GPa ($KIH_2$). Especially, the commonly known table salt, NaCl, can form stable $NaClH_2$ and $NaCl(H_2)_4$ hybrid compounds under pressures of 19.7 and 37.9 GPa, respectively. Our results demonstrate that pressure can promote the formation of solid-molecule hybrid compounds, which reveals a new unique way of discovering novel materials that can combine the advantages of inorganic compounds and small molecules. A direct application of our work is the proof of turning the common table salt and other similar ionic compounds into hydrogen storage materials under moderate pressures.

Hybrid materials merge the organic and inorganic components in the same structure and therefore might combine the strength of the two[1]. However, despite the long-time effort, most of the hybrid materials are nanocomposites and the true examples of compounds that consists of molecules that embedded in the inorganic solid crystals at the atomic level is very rare. Recently the astonishing application of hybrid perovskite materials in converting solar energy into electricity with high efficiency caught great attention. These materials consist of p block metals and a late halogen such as Pb and I[2-4]. The third component can be an organic molecule such as $CH_3NH_3$[5,6]. The superior properties and the easy and low-cost processing of these materials encourage the search of new hybrid materials. However, the lack of chemical interactions between the inorganic and the organic components and the large differences between the geometry and electronic structures of the two kinds prevent the formation of the stable compounds.

Dissimilar to ambient condition, materials might behave very differently under high pressure. Especially, as shown by many examples in recent studies, pressure can stabilize many novel compounds. Some elements, such as Li and Be, that do not form stable compounds under ambient condition can react under high pressure[7]. Compounds with very unusual compositions such as $Na_3Cl$ can form under high pressure[8]. More strikingly, the most stable element He can form stable compounds with many ionic compounds under high pressure (such as, $Na_2He$, $MgF_2He$, and $FeO_2He$, etc.)[9-11]. These He insertion reactions show very large formation energies, although there are apparently no strong chemical bonds forming between He and the neighboring atoms.

Hydrogen gas molecules are similar to helium in volume and the chemical reactivity under standard conditions, and therefore might behave similarly as He under high pressure.

Hydrogen is the simplest and most abundant element in the universe. It has the potential to become an efficient fuel due to its clean and environmental friendly products and high gravimetric energy density. The technology of using hydrogen as a fuel remains as an active research area worldwide[12,13] for decades. One major challenge is the storage of hydrogen. Most commonly, hydrogen is stored either in high pressure tanks or in liquid form in cryogenic tanks[14,15]. These forms of storage are not suitable for widespread commercial application due to their low energy density, high costs, and safety issues. As an alternative, solid materials for hydrogen storage have been persued including hydrogen adsorption materials[16-21], clathrate hydrates[22], host metals that can adsorb hydrogen in their interstitial sites[23], chemically bonded covalent and ionic compounds and some reactive metals[24] since they can improve the hydrogen density effectively. Despite all the above progresses, the search of simple, low-cost and easy processing hydrogen storage materials with superb hydrogen adsorption and release kinetics remains as a challenging key issue in the area of hydrogen fuel.

In this work, we examine the possibility of forming hybrid materials by combining small molecules and inorganic compounds. We choose alkali halides as the representing ionic compounds and $H_2$ as the example molecules. We study the stability of possible XY-$H_2$ compounds with various compositions under high pressures up to 100 GPa (Fig. 1 and S1-8), using density functional method and the crystal structure search techniques.

Strikingly, we found that many alkali halides can form stable compounds with $H_2$, some under a pressure as low as 0.8 GPa. Especially, NaCl, the most common ionic compound known as the table salt, can form $NaClH_2$ and $NaCl(H_2)_4$ under the pressures of 19.7 and 37.9 GPa.

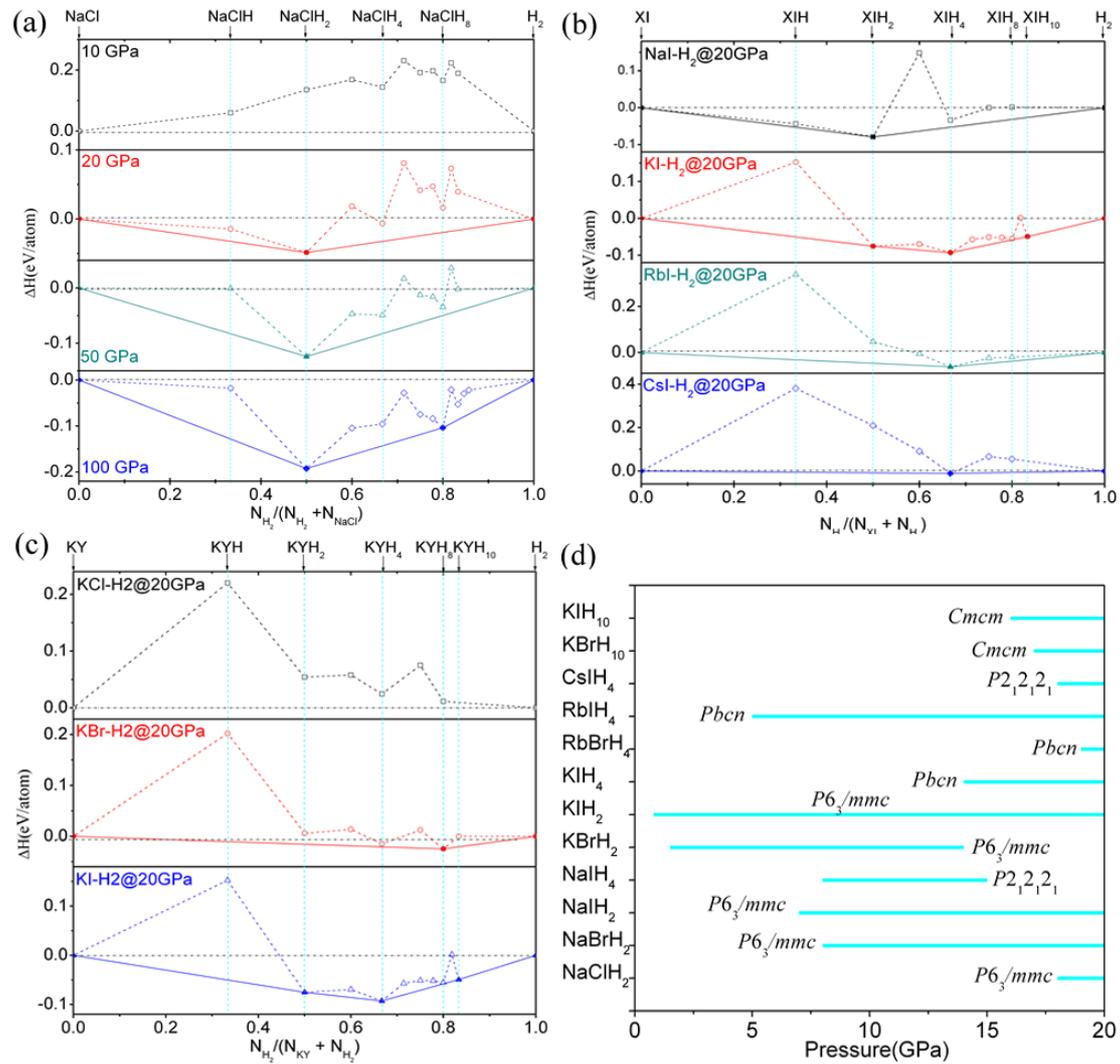

**Figure 1. Phase stabilities of various H-rich XY-$H_2$ hybrid compounds**. Enthalpies of formation of (a) NaCl-$H_2$ under several pressures up to 100 GPa; (b) XY-$H_2$ (X=Na, K, Rb, Cs; Y=I) under 20 GPa; (c) XY-$H_2$ (X=K; Y=Cl, Br, I) under 20 GPa. Dotted lines connect the data points, and solid lines denote the convex hull. (d) Predicted pressure-composition phase diagram of XY salt hydrides.

The thermodynamic stability of XY-$H_2$ hybrid compounds are evaluated from their formation enthalpies relative to the dissociation products XY + $H_2$. In principle, XY-$H_2$ are ternary compounds and may have many different possible decomposition products. However, all of our structure searches show no sign of $H_2$ dissociation, therefore $H_2$ could be treated as a reaction unit throughout our study. On the other hand, XY is the most stable stoichiometry for most of the metal halides at ambient condition and the high-pressure range (<100 GPa) in this study. Therefore, we reduce the stability evaluation to a binary reaction of XY+n$H_2$ → XY($H_2$)$_n$.

The stability of the XY-$H_2$ compounds strongly depend on the alkali metals, the halogens and the pressure. As shown in Fig. 1a, the increase of the pressure will significantly improve the stability of NaCl$H_2$. NaCl$H_2$ becomes stable at a pressure of 19.7 GPa. While pressure changes from 20 GPa to 100 GPa, the formation enthalpy of NaCl$H_2$ improves from -0.049 eV/atom to -0.192 eV/atom. Furthermore, pressure can also promote the formation of stable XY-$H_2$ compounds with higher $H_2$ compositions. At 20 GPa, NaCl$H_2$ is the only stable composition. While at 100 GPa, both NaCl$H_2$ and NaCl($H_2$)$_4$ (containing 4 $H_2$ molecules) become stable. Fig. 1b show the convex hulls of iodide-$H_2$ compounds. Interestingly, the compounds become less stable with increasing cation radius from NaCl to CsCl. Fig. 1c shows the stability of hybrid compounds formed by potassium halides. In contrast to the trend of cation, the increasing of the anion size improves the stability of the hybrid compounds. The above results reveal a general trend that the differences of the sizes of the cation and the anion is a determinate factor of forming stable hybrid compounds with $H_2$ molecules. That is why KI form hybrid compound with $H_2$ at the lowest pressure of 0.8 GPa. Furthermore, more $H_2$ molecules can be incorporated for ionic compounds with larger size difference. For example, three compositions can form for KI-$H_2$ compounds with 1, 2, and 5 $H_2$ molecules in one KI formula at 20 GPa. The stable pressure range and the most stable structures of studied XY-$H_2$ hybrid compounds are summarized in Fig. 1d.

The high-pressure behavior of NaCl has been extensively studied experimentally at pressures up to 304 GPa[25-28] and by ab initio simulations[29-32], and very simple behavior was observed: At 30 GPa, the rocksalt structure was found to transform into the CsCl (B2-type) structure[28,30]. From the Herzfeld criterion, metallization of NaCl is expected to occur at 300 GPa[32], whereas density functional theory calculations[31] suggest 584 GPa. In addition, the unexpected stoichiometries of Na-Cl system can be stable in the environment of Na-rich or Cl-rich[8]. Recently, $Na_2Cl$ and $Na_3Cl$ together with regular NaCl as two-dimensional Na–Cl crystals, is observed under ambient conditions[33]. However, due to the strong ionic interactions, only NaCl is the decomposition product of NaClH$_n$.

The above results already included the effects of van der Waals (vdW) interactions and the zero-point energies. These effects are important and can change the transition pressures but not crucial for the formation of XY-$H_2$ hybrid compounds. The test calculations show that the phase transition pressure of *Pc* to *P*6$_3$/*mmc* is 14.6 GPa comparing the normal PBE result of 16.5 GPa for NaClH$_2$ (Fig. S27), and the formation pressures of NaClH$_2$ are 15.3 and 17.4 GPa with and without vdW interactions, respectively. While the ZPE is added, the formation pressure for NaClH$_2$ becomes 19.7 GPa.

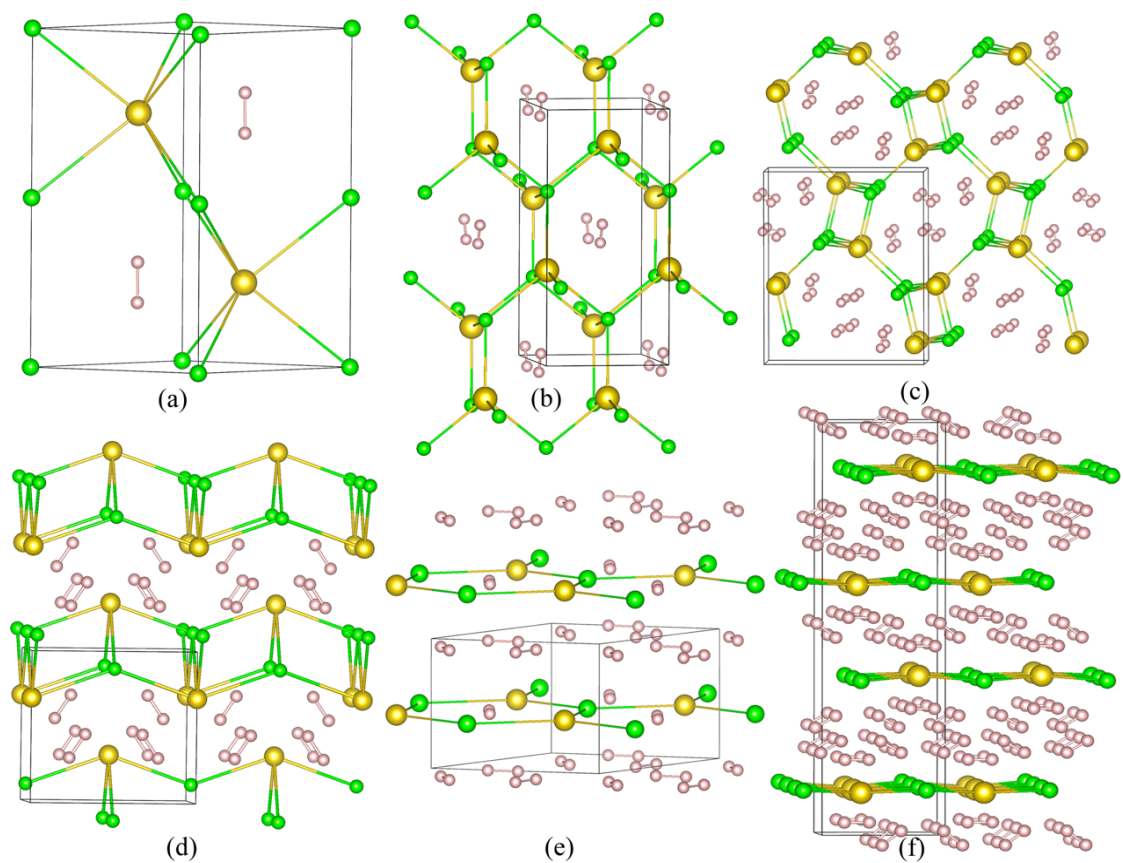

**Figure 2. Structures of XYH$_n$ compounds.** Three different ways of inserting H$_2$ molecules inside the lattices of the XY ions: at atomic sites in $P6_3/mmc$ symmetry (a); inside tubes in XYH$_2$ with $Pc$ symmetry (b) and XY(H$_2$)$_2$ with $P2_12_12_1$ symmetry (c); and between layers in XY(H$_2$)$_2$ with $Pbcn$ symmetry (d), XY(H$_2$)$_4$ with $Pm$ symmetry (e), and XY(H$_2$)$_5$ with $Cmcm$ symmetry (f). The white, gold and green spheres represent H, alkali and halogen atoms, respectively.

The PSO structure search reveals unique and surprisingly simple structure features for the XY-H$_2$ hybrid compounds. All XYH$_2$ compounds are found to adopt a hexagonal structure with $P6_3/mmc$ symmetry. Most of the XY(H$_2$)$_2$ compounds except NaI(H$_2$)$_2$ and CsI(H$_2$)$_2$ adopt a $Pbcn$ structure. The two exceptions, NaI(H$_2$)$_2$ and CsI(H$_2$)$_2$, adopt the $P2_12_12_1$ structure. These structures reveal three different ways of inserting H$_2$ molecules inside the lattices of the XY ions: at atomic sites (Fig. 2a), inside tubes (Fig.

2b and 2c) and between layers (Fig. 2d-2f). In the $P6_3/mmc$ structure for the $XYH_2$ hybrid compounds, the sub-lattice of XY is isomorphous to the NiAs structure, in which X and Y ions occupy 2c (1/3, 2/3, 1/4) and 2a (0, 0, 0) Wyckoff positions, respectively. The centroid of $H_2$ molecule occupies the 2d (1/3, 2/3, 3/4) positions. While the hydrogen composition increases, the $H_2$ molecules start to agglomerate. In $XY(H_2)_2$ compounds ($P2_12_12_1$ structure, Fig. 2c), $H_2$ molecules are inserted in the tubes that are formed by X and Y ions. Similar inserted tube structure is also found for a $NaClH_2$ metastable structure below 14.6 GPa. The *Pbcn* structure (Fig. 2d) for $XY(H_2)_2$ hybrid compounds consist of double layers of bended XY sheets, in which both metal and halogen ions are 4-fold coordinated. Taking $RbI(H_2)_2$ as an example, the shortest Rb-I distance within a layer is 3.65 Å at ambient condition, which is comparable to the Rb-I distance of 3.7 Å in RbI compound in the rocksalt structure (*Fm*-3*m*). In contrast, the Rb-I distances between the layers is much larger (4.61 Å). Hydrogen present in the form of $H_2$ molecules inserted in between the layers. The bond length of $H_2$ molecules is 0.72 Å while the structure is relaxed at 0 GPa, which is very close to the bond length of $H_2$ in the molecular phase *C*2/*c* at ambient condition.

In hybrid compounds with higher $H_2$ composition, $H_2$ molecules could occupy both atomic sites and the space between layers such as $XY(H_2)_4$ in *Pm* structure, or are inserted in between the single layers of XY layers such as $XY(H_2)_5$ in *Cmcm* structure. $NaCl(H_2)_4$ is found to become stable above 37.9 GPa and adopts a structure with *Pm* symmetry (Fig. 2d). Above 78 GPa, the orthorhombic structure with *Cmcm* symmetry (Fig. S28) is more stable than the *Pm* phase. These structures are demonstrated to be dynamically stable in their stable pressure ranges by phonon calculations (Fig. 3d and S24). In the *Cmcm* structure, Na and Cl atoms with coordination number of 4, form the NaCl sheet with the shortest distance of Na-Cl is 2.31 Å at 100 GPa, which is much shorter than that (2.7 Å) of *Fm-3m* NaCl at ambient condition, and the distance of between the nearest NaCl sheets is 3.28 Å. There are four $H_2$ molecules between the two NaCl sheets with the bond length of $H_2$ molecule is 0.72 Å at 150 GPa, which nearly equals to that of solid $H_2$ at ambient pressure. Comparing to NaCl$(H_2)_5$, KI$(H_2)_5$ becomes stable at a lower pressure of 17 GPa (Fig. 1b) and adopts the same *Cmcm* ctructure. The shortest distance of K-I is 3.17 Å at 20 GPa, which is much shorter than that (3.19 Å) of *Fm-3m* KI at 20 GPa, and the distances of between the nearest KI sheets are 4.97 and 4.69 Å. Three and two $H_2$ molecules are located between the nearest KI sheets with the shorter and longer distance. In which, the bond length of $H_2$ molecule is 0.72 Å at 20 GPa.

Although XY-$H_2$ hybrid compounds are only stable under high pressure, most of them are found to be recoverable after the pressure is partially or completely released. For example, the phonon calculations reveal no imaginary vibrational modes for the NaClH$_2$ in *Pc* phase (Fig. 3a), KIH$_2$ in the *P6$_3$/mmc* structure (Fig. 3b), and RbI$(H_2)_2$ in *Pbcn* structure (Fig. 3c) at ambient condition. Therefore, these new type of hybrid compounds have the potential to be used as effective hydrogen storage materials. Although their volumetric and gravimetric densities are not very high, they have the potential to become an easy processing and low-cost hydrogen storage materials. More interestingly, we found that in some hybrid compounds, the structure of the XY sublattice might remain metastable after remove all the hydrogen molecules. For example, the NaCl lattice in the *Pc* phase of NaClH$_2$ remains in the tube-like structure after the removal of hydrogen molecules at ambient condition, forming a novel phase of NaCl with the coordinate number of 5. Phonon calculations (Fig. S26) reveals no imaginary vibrational modes for this novel structure of NaCl. This structure has a gap of 4.6 eV that is significantly smaller than that of rocksalt NaCl.

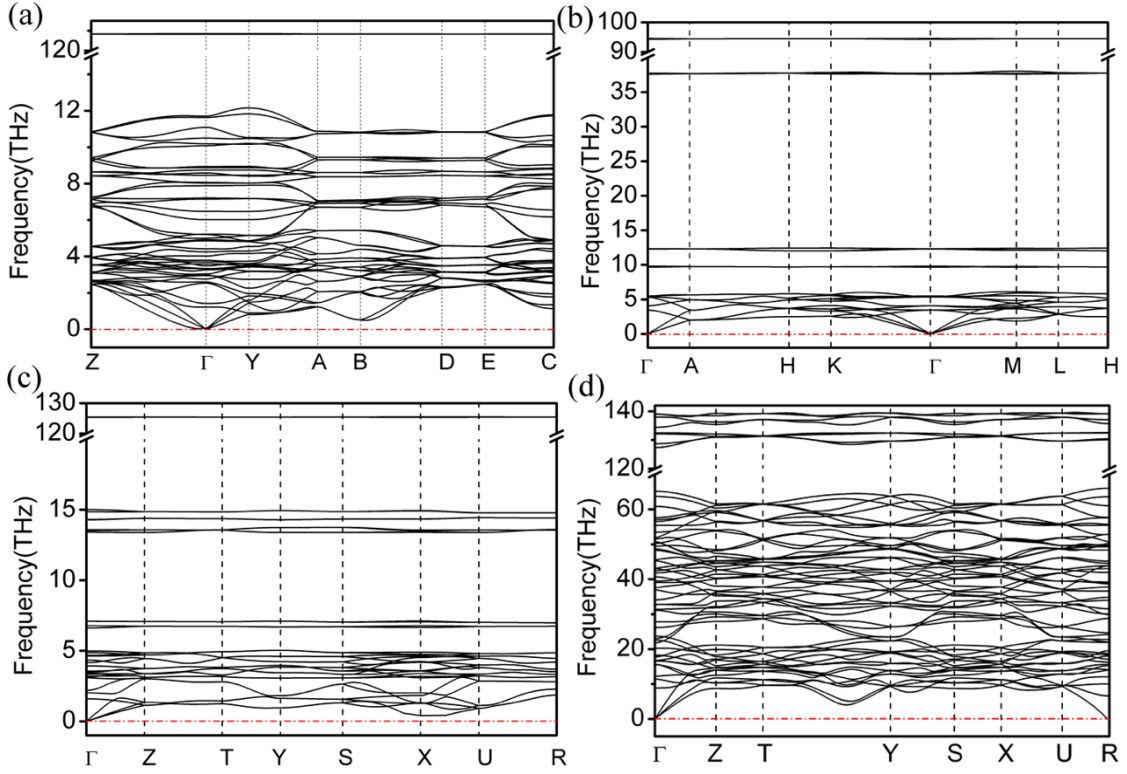

**Figure 3. The phonon spectrum for XY-H$_2$ hybrid compounds**. The phonon spectrum of *Pc* phase for NaClH$_2$ (a) at zero pressure, *P*6$_3$/*mmc* phase for KIH$_2$ (b) zero pressure, *Pbcn* phase for RbI(H$_2$)$_2$ (c) at zero pressure, and *Cmcm* phase for NaCl(H$_2$)$_4$ at 100 GPa.

To probe the thermodynamic origin of the stabilization of the XY-H$_2$ hybrid compounds, we chose the NaCl-H$_2$ compounds as example systems and split their reaction enthalpies ΔH into the internal energy changes ΔU and the pressure-volume terms ΔPV. For NaClH$_2$, although ΔU is positive at low pressure whereas ΔPV is negative, both terms decrease with increasing pressure, contributing to the continuous decrease of ΔH and therefore to the stabilization of the hybrid compound. Especially, the ΔPV term decreases much faster than ΔU, indicating that the formation of the XY-

$H_2$ hybrid compounds is mainly due to the volume reduction. For compounds with high H composition, such as $NaCl(H_2)_4$, $\Delta PV$ has a very low negative value that changes only slightly with the pressure. It is the significant decrease of the $\Delta U$ that eventually stabilize these hybrid compounds.

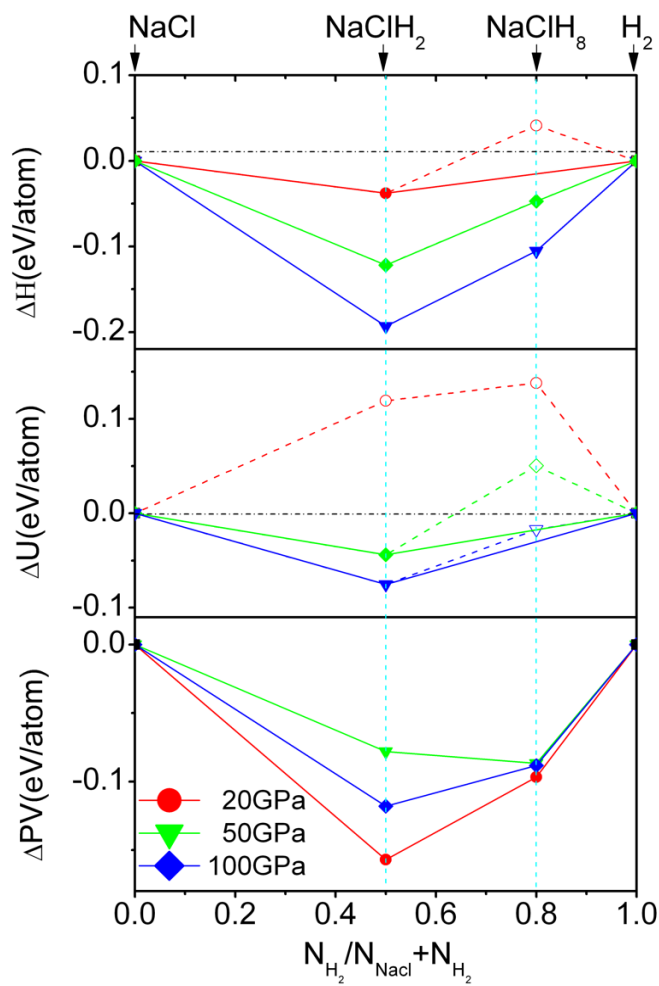

**Figure 4. Energy contributions to the formation of NaCl-H2 hybrid compounds.** Reactions enthalpies (top panel), internal energy changes (middle panel), and PV term changes (bottom panel) for the formation of $NaCl-H_2$ hybrid compounds at different pressures.

Furthermore, we examine the electronic structures of the XY-$H_2$ hybrid compounds and found that the inserted $H_2$ molecules have no considerable chemical interactions with the ionic sublattices. The first evidence is from the band structure and density of states. The results for NaCl$H_2$ in *Pc* and *P*$6_3$/*mmc* structures, KI$H_2$ in *P*$6_3$/*mmc* structure, and RbI($H_2$)$_2$ in *Pbcn* structure at ambient pressure are presented in Fig. 3d and S12. They show that the bands in a large energy range around the Fermi level are dominated by the states from XY sublattices. Large band gaps of nearly 5.6, 5.7, 5.3 and 4.3 eV were found in these structures. In order to test the effect of $H_2$ insertion to the electronic structure of the ionic sublattice, we constructed a model system, NaCl$H_0$ in which all the $H_2$ molecules are removed from the *Pc* (Fig. 5a) and *P*$6_3$/*mmc* (Fig. S9) structures. The results show that the $H_2$ molecules do not have significant effects to change the bands around the Fermi energy, indicating there is no strong chemical interactions between $H_2$ molecules and the ionic XY sublattice. Using Bader charge analysis (Table S1), we found that valence electrons of Na atoms transfer almost completely Cl atoms, while the $H_2$ molecules don't obtain charges from Na. This is consistent to the fact that the bond length of $H_2$ in XY($H_2$)$_n$ is similar to that in the pure solid $H_2$.

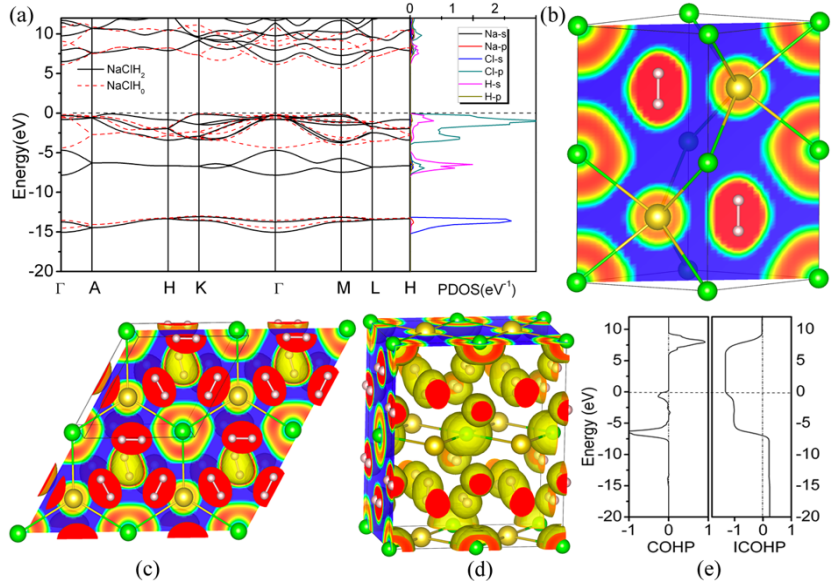

**Figure 5. Calculated electronic properties for NaClH$_n$ at various pressures**. (a) The electronic band structure and PDOS of *Pc* phase for NaClH$_2$ at 0 GPa. In the left panel, the black solid lines are electronic band structure of NaClH$_2$; the red dashed lines are those of NaClH$_0$ in which all the H$_2$ molecules are removed from the NaClH$_2$ structure. The black and red dashed lines show the Fermi energy of NaClH$_2$ and NaClH$_0$. The right panel presents the projected DOS of NaClH$_2$. (b) – (d) The calculated ELFs of NaClH$_2$ with *P*6$_3$/*mmc* phase at 20 GPa, NaCl(H$_2$)$_4$ with *Pm* symmetry at 50 GPa, and NaCl(H$_2$)$_4$ with *Cmcm* symmetry at 100 GPa. (e) Calculated COHP and ICHOP of NaClH$_2$ at 50 GPa.

We also directly examine the bonding feature of XY-H$_2$ hybrid compounds. First, taking NaCl-H$_2$ compounds as examples, we calculated their electron localization functions (ELF)[34] (Figs. 5b – 5d) under various pressures. For NaClH$_2$, the nature of the Na-Cl bonds is purely ionic in view of the absence of charge localization between Na and Cl (Fig. 5b), while the H-H bonds are strongly covalent as shown by the large values of the ELF between the nearest-neighbor H atoms. The ELFs of *Pm* phase at 50 GPa (Fig. 5c) and *Cmcm* phases for NaCl(H$_2$)$_4$ at 100 GPa (Fig. 5d) show the same bonding nature between Na-Cl and between H-H, i.e. the former is ionic and the latter is covalent. There is no strong bonding between H$_2$ molecules and the surrounding ions for both compounds. Similar results are found for all other XY-H$_2$ hybrid compounds, revealing they share very similar bonding features.

To examine the covalent bonds in the compounds, the crystal orbit Hamiltonian population (COHP)[35] method is used, which counts the population of wavefunctions on two atomic orbitals of a pair of selected atoms. The calculated COHP and integrated COHP (ICOHP) clearly show that the H-H bonding states are occupied whereas the antibonding states are not, supporting the strong covalent nature of H-H (Fig. 5e). The integrated COHP (ICOHP) can provide an estimate of the strength of bonding. For comparison, we have calculated the ICOHP values up to the Fermi level for H-H pairs in $XYH_2$-type salt hydrides and solid hydrogen at various pressures. The ICOHP value with -1.3 pairs/eV of H-H bond reveal that the strength of H-H in $XYH_2$-type salt hydrides increases with the higher pressure.

**Conclusions**

In conclusion, we propose that high pressure can be used as a unique method to synthesize hybrid materials that contain both inorganic sublattices and inserted molecules. Using DFT calculations and crystal structure search methods, we show that alkali halides (XY) can react with $H_2$ under moderate pressures as low as 0.8 GPa and form stable $XY(H_2)_n$ hybrid compounds. Hydrogen maintains its molecular form in all phases. Depending on the pressure and the sizes of the ions, different number of $H_2$ molecules can be inserted at atomic sites, inside the tubes or between the layers of XY sublattices. Most of the hybrid compounds are found to be recoverable after the release of the external pressure. Especially, $H_2$ can react with NaCl, the commonly known table salt, and form $NaClH_2$ and $NaCl(H_2)_4$ at the pressures of 19.7 and 37.9 GPa. The electronic structure calculations show that $H_2$ molecules do not form any strong local chemical bond with the surrounding ions. It is mainly the reduction of the volume that drives the formation of these compounds under high pressure. These results show a new path toward synthesizing hybrid materials that may combine the strength and unique properties of both inorganic solid compounds and the small molecules. It also shows that simple ionic compounds can be turned into hydrogen storage materials under pressure.

Many hydrogen rich materials including various hydrides and superhydrides have been obtained by high pressure synthesis techniques in the last decade. Some of them, such as $H_3S$[36], $LaH_{10}$ and $YH_{10}$[37] show transitions to superconducting states with a $Tc$ approaching room temperature. The bonding nature of these compounds is very different to that of the XY-$H_2$ hybrid compounds. In those hydrides, there are large charge transfer between hydrogen atoms and the surrounding atoms, causing the dissociation of the $H_2$ molecules. On the other hand, there are many metal hydrides, such as $NaH_n$[38], $LiH_n$[39] etc, were predicted to contain $H_2$ molecules in their crystal lattices. As a matter of fact, some of the hydrides can be viewed as XH-$H_2$ hybrid compounds. Our work reveals that the structure features hidden in these compounds with unusual compositions might have much broader significance and potential applications.

**Methods**

**Crystal structure prediction.** Our structure searching simulations are performed through the swarm-intelligence based CALYPSO method[40] via a global minimization of free energy surfaces merging *ab initio* total-energy calculations as implemented in the CALYPSO code[41]. The method is specially designed for global structural minimization unbiased by any known structural information, and has been benchmarked on various known systems[37,42,43].

**Total energy calculations.** Total energy calculations were performed in the framework of density functional theory within the Perdew-Burke-Ernzerhof[44] parameterization of generalized gradient approximation[45] as implemented in the VASP (Vienna Ab Initio simulation package) code[46]. The projector-augmented wave (PAW) method[47] was adopted with the PAW potentials taken from the VASP library where d electrons are treated as valence electrons for alkali elements. The use of the plane-wave kinetic energy cutoff of 1200 eV and dense k-point sampling, adopted here, were shown to give excellent convergence of total energies. We explored the zero point energy effects on the formation energy using the phonopy code[48].

**Author Contributions** M. M. proposed the research. F.P. performed the calculations and processed data. F. P. and M. M. analyzed and interpreted the results. M. M., F. P., and Y. M. wrote the paper.

**Author Information** Correspondence and request for materials should be addressed to M. M. (mmiao@csun.edu).

**Acknowledgements** M.M. acknowledges the support of NSF CAREER award 1848141 and ACF PRF 50249-UN16. F.P. and Y.M. acknowledge funding support from the National Natural Science Foundation of China under Grant No. 11774140, China

Postdoctoral Science Foundation under Grant No. 2016M590033, the Natural Science Foundation of Henan Province under Grant No. 162300410199, Program for Science and Technology Innovation Talents in University of Henan Province under Grant No. 17HASTIT015, and Open Project of the State Key Laboratory of Superhard Materials, Jilin University under Grant No. 201602.